\documentclass[a4paper,10pt]{article}

\usepackage[top=1.5in, bottom=1.5in, left=1in, right=1in]{geometry}

\usepackage{graphicx}
\usepackage{mathrsfs}
\usepackage{mathtext}
\usepackage{amsmath}

\title{Parameter Appendix}
\author{Bo Tan, John Thompson}

\begin{document}

\maketitle

This is a support document which describes the properties of the cable and parameters of the formulas in \cite{Formal_Paper}. The cable parameters help the reader build powerline channel according to the transmission line theory. The document also presents the parameters which describe the distribution of the number of path, path magnitude, path interval and the cable loss feature of the powerline channel. By using the parameters in this document, readers can model the powerline channel according to the methodology proposed in \cite{Formal_Paper}. 

\section{Cable Property}
Figure. 1 shows the structure of a power cable which is used in \cite{Multi_path_Zimmermann}, Table 1 and Table 2 show the geometric parameters and the electromagnetic parameters of NAYY150 and NAYY35 cables. The insulator between conductors is PVC. When feeding signals into two adjacent conductors, most of the electric field is concentrated between these two conductors. The lumped parameters of the cable can be calculated by the geometric dimensions and material electrical properties.
\begin{figure}[h]
\begin{center}
 \includegraphics[scale=.5]{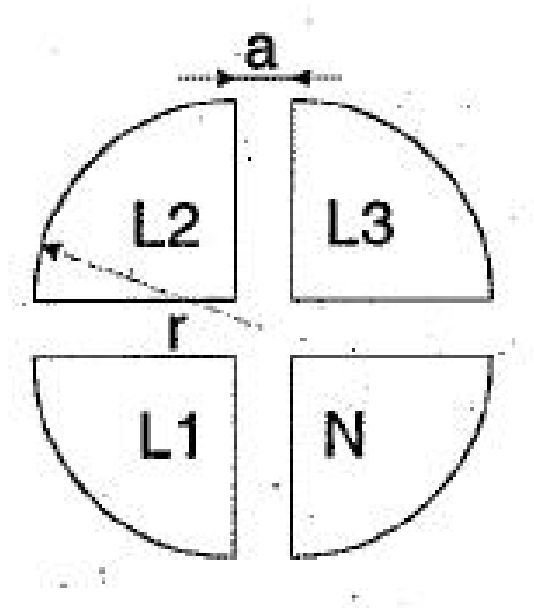}
 % section.eps: 0x0 pixel, 300dpi, 0.00x0.00 cm, bb=0 668 166 842
  \caption{The Cross-Section of a Four Conductor Power Cable, with three live connections (\textsf{L1}, \textsf{L2} and \textsf{L3}) and one neutral connection (\textsf{N}). The scalar $a$ represents the space between two adjacent conductors which is filled by PVC. Scalar $r$ is the radius from geometry center to the outer edge of the conductor.}
\end{center}
\end{figure}

\begin {table}[h]
\caption {Cable Geometry Properties}
\begin{center} 
\begin{tabular}{|p{0.25in}|p{1.5in}|p{1.5in}|}
\hline
  & NAYY150 (mm) & NAYY35 (mm) \\ \hline
a  & 1.8 & 1.2 \\ \hline
r  & 6.9099 & 5.9161 \\ \hline
\end{tabular}
\end{center}
\end {table}

\begin {table}[h]
\caption {Cable Electrical Properties}
\begin{center}
% use packages: array
\begin{tabular}{|l|l|l|} 
\hline
Conductivity of Cooper & $\kappa$ & $58\times 10^{6}$ S/m \\ \hline
Dissipation of PVC & tan$\delta$ & 0.025 \\ \hline
Relative permittivity of PVC & $\varepsilon_r$ & 4  \\ \hline
Free space permittivity & $\varepsilon_0$ & $8.5419\times 10^{-12}$ F/m  \\ \hline
Relative permeability of Cooper & $\mu_r$ & 1  \\ \hline
Free space permeability & $\mu_0$ & $1.2566 \times 10^{-6}$ μ [H/m]   \\ \hline
\end{tabular}
\end{center}
\end{table}

The lumped parameters such as capacitance ($C$), inductance ($L$), resistance ($R$) and conductance ($G$) per unit length can be calculated by applying the parameters in Table 1 and 2 to equation (1) to (4).
\begin{eqnarray} \label{eq:Appendix_C}
  C&=&\varepsilon_{0}\varepsilon_{r}\dfrac{r}{a} \\
  L&=&\mu_{0}\mu_{r}\dfrac{r}{a} \\
  R&=&\sqrt{\dfrac{\pi \mu_{0}}{\kappa r^{2}} f} \\
  G&=&2\pi fC \textrm{tan}\delta
\end{eqnarray}

%\begin{equation} \label{eq:Appendix_L}
%  L=\mu_{0}\mu_{r}\dfrac{r}{a}
%\end{equation}

%\begin{equation} \label{eq:Appendix_R}
%  R=\sqrt{\dfrac{\pi \mu_{0}}{\kappa r^{2}} f}
%\end{equation}

%\begin{equation} \label{eq:Appendix_G}
%  G=2\pi fC \textrm{tan}\delta
%\end{equation}

where, $f$ in equation (3) and (4) denotes the frequency with Hz as unit.

\section{Parameters for Number of Paths Distribution}
The number of the paths for the channel of the $i$th Class and the $k$th Cluster can be described by equation (7) in \cite{Formal_Paper} which is a Gaussian distribution:
\begin{equation}
 N_{i,k}=\left[ \mathcal{N} (\mu_{i,k},\sigma^{2}_{i,k}) \right] 
\end{equation} 
where $\left[\cdot \right] $ means to round towards the nearest integer, parameters $\mu_{i,k}$ and $\sigma^{2}_{i,k}$ are the expectation and standard deviation of the Gaussian distribution. The value $\mu$ and $\sigma$ in each Class increase as power function of Cluster Index. The Power function can be written as:
\begin{equation}
 \mu_{i,k}=p_{i1}k^{p_{i2}}+p_{i3}
\end{equation} 
\begin{equation}
 \sigma^{2}_{i,k}=q_{i1}k^{q_{i2}}+q_{i3}
\end{equation}
The parameters $p_{i1}$, $p_{i2}$, $p_{i3}$, $q_{i1}$, $q_{i2}$ and $q_{i3}$ are shown in Table 3:
\begin {table}
\caption {Parameters for the Path Number distributions}
\begin{center}
\begin{tabular}{|l|lll|}
\hline
i=2 & $p_{i1}=1.623$ & $p_{i2}=0.08596$ & $p_{i3}=0$ \\
    & $q_{i1}=-0.3818$ & $q_{i2}=-0.8461$ & $q_{i3}=1.592$ \\ \hline

i=3 & $p_{i1}=3.913$ & $p_{i2}=0.0968$ & $p_{i3}=0$ \\
    & $q_{i1}=-0.005983$ & $q_{i2}=1.033$ & $q_{i3}=1.783$ \\ \hline

i=4 & $p_{i1}=6.169$ & $p_{i2}=0.09686$ & $p_{i3}=0$ \\
    & $q_{i1}=-0.4401$ & $q_{i2}=-0.6989$ & $q_{i3}=2.007$ \\ \hline

i=5 & $p_{i1}=8.684$ & $p_{i2}=0.1688$ & $p_{i3}=0$ \\
    & $q_{i1}=-8.725$ & $q_{i2}=-0.06764$ & $q_{i3}=10.98$ \\ \hline
\end{tabular}  
\end{center}
\end{table}

\section{Parameters for Magnitude Distribution}
\subsection{First Arrival Path}
For the channels in a particular Class, the magnitude of the first arrival path of the $i$th Class follows a double exponential decay distribution with the increase of the Cluster Index $k$. In \cite{Formal_Paper}, equation (8) is used to describe the double exponential decay which can be written as:
\begin{equation}
 I_{i,k}=a_{i}^{M} e^{b_{i}^{M}k}+c_{i}^{M}e^{d_{i}^{M}k}
\end{equation} 
where $a_{i}^{M}$, $b_{i}^{M}$, $c_{i}^{M}$ and $d_{i}^{M}$ are the double exponential parameters, and are shown in Table 4:
\begin {table}
\caption {Parameters for the First Arrival Path Magnitude distribution}
\begin{center}
\begin{tabular}{|l|llll|}
\hline
i=1 & $a_{i}^{M}=0.4815$ & $b_{i}^{M}=-0.0821$ & $c_{i}^{M}=0.4103$ & $d_{i}^{M}=-0.02408$ \\ \hline
i=2 & $a_{i}^{M}=0.2601$ & $b_{i}^{M}=-0.1214$ & $c_{i}^{M}=0.4948$ & $d_{i}^{M}=-0.03241$ \\ \hline
i=3 & $a_{i}^{M}=0.1841$ & $b_{i}^{M}=-0.1246$ & $c_{i}^{M}=0.3628$ & $d_{i}^{M}=-0.03334$ \\ \hline
i=4 & $a_{i}^{M}=0.1221$ & $b_{i}^{M}=-0.1515$ & $c_{i}^{M}=0.2736$ & $d_{i}^{M}=-0.03445$ \\ \hline
i=5 & $a_{i}^{M}=0.1721$ & $b_{i}^{M}=-0.1517$ & $c_{i}^{M}=0.0905$ & $d_{i}^{M}=-0.01979$ \\ \hline
\end{tabular}  
\end{center}
\end{table}

\subsection{Other paths}
The magnitude of the other paths are dependent on the arrival time of the path. The decay profile with the time sampling index can be also demonstrated by the double exponential function which is also can be seen in \cite{Formal_Paper} as equation (9): 
\begin{equation}
 I_{k,j}=a_{k}^{o} e^{b_{k}^oj}+c_{k}^{o}e^{d_{k}^{o}j}
\end{equation}
where $k$ is the cluster index. $a_{i}^{o}$, $b_{i}^{o}$, $c_{i}^{o}$ and $d_{i}^{o}$ are the double exponential parameters, and are shown in Table 5:
\begin {table}
\caption {Parameters for the Other Path Magnitude distribution}
\begin{center}  
\begin{tabular}{|l|llll|}
\hline
k=1 & $a_{k}^{o}=0.4194$ & $b_{k}^{o}=-0.1270$ & $c_{k}^{o}=0.0328$ & $d_{k}^{o}=-0.0083$ \\ \hline
k=2 & $a_{k}^{o}=0.4388$ & $b_{k}^{o}=-0.1355$ & $c_{k}^{o}=0.0487$ & $d_{k}^{o}=-0.0207$ \\ \hline
k=3 & $a_{k}^{o}=0.4647$ & $b_{k}^{o}=-0.1353$ & $c_{k}^{o}=0.0502$ & $d_{k}^{o}=-0.0206$ \\ \hline
k=4 & $a_{k}^{o}=0.4542$ & $b_{k}^{o}=-0.1329$ & $c_{k}^{o}=0.0562$ & $d_{k}^{o}=-0.0235$ \\ \hline
k=5 & $a_{k}^{o}=0.4381$ & $b_{k}^{o}=-0.1244$ & $c_{k}^{o}=0.0521$ & $d_{k}^{o}=-0.0229$ \\ \hline
k=6 & $a_{k}^{o}=0.4632$ & $b_{k}^{o}=-0.1253$ & $c_{k}^{o}=0.0571$ & $d_{k}^{o}=-0.0249$ \\ \hline
k=7 & $a_{k}^{o}=0.4677$ & $b_{k}^{o}=-0.1163$ & $c_{k}^{o}=0.0422$ & $d_{k}^{o}=-0.0196$ \\ \hline
k=8 & $a_{k}^{o}=0.5124$ & $b_{k}^{o}=-0.1200$ & $c_{k}^{o}=0.0457$ & $d_{k}^{o}=-0.0213$ \\ \hline
k=9 & $a_{k}^{o}=0.4262$ & $b_{k}^{o}=-0.1032$ & $c_{k}^{o}=0.0327$ & $d_{k}^{o}=-0.0171$ \\ \hline
k=10 & $a_{k}^{o}=0.4419$ & $b_{k}^{o}=-0.1004$ & $c_{k}^{o}=0.0287$ & $d_{k}^{o}=-0.0151$ \\ \hline
k=11 & $a_{k}^{o}=0.5116$ & $b_{k}^{o}=-0.1046$ & $c_{k}^{o}=0.0292$ & $d_{k}^{o}=-0.0149$ \\ \hline
k=12 & $a_{k}^{o}=0.4604$ & $b_{k}^{o}=-0.0964$ & $c_{k}^{o}=0.0257$ & $d_{k}^{o}=-0.0140$ \\ \hline
k=13 & $a_{k}^{o}=0.4501$ & $b_{k}^{o}=-0.0925$ & $c_{k}^{o}=0.0223$ & $d_{k}^{o}=-0.0126$ \\ \hline
k=14 & $a_{k}^{o}=0.4968$ & $b_{k}^{o}=-0.0946$ & $c_{k}^{o}=0.0238$ & $d_{k}^{o}=-0.0134$ \\ \hline
k=15 & $a_{k}^{o}=0.5187$ & $b_{k}^{o}=-0.0950$ & $c_{k}^{o}=0.0243$ & $d_{k}^{o}=-0.0136$ \\ \hline
k=16 & $a_{k}^{o}=0.5242$ & $b_{k}^{o}=-0.0915$ & $c_{k}^{o}=0.0207$ & $d_{k}^{o}=-0.0116$ \\ \hline
k=17 & $a_{k}^{o}=0.5355$ & $b_{k}^{o}=-0.0896$ & $c_{k}^{o}=0.0188$ & $d_{k}^{o}=-0.0109$ \\ \hline
k=18 & $a_{k}^{o}=0.6164$ & $b_{k}^{o}=-0.0934$ & $c_{k}^{o}=0.0224$ & $d_{k}^{o}=-0.0125$ \\ \hline
k=19 & $a_{k}^{o}=0.5288$ & $b_{k}^{o}=-0.0852$ & $c_{k}^{o}=0.0180$ & $d_{k}^{o}=-0.0108$ \\ \hline
k=20 & $a_{k}^{o}=0.5829$ & $b_{k}^{o}=-0.0864$ & $c_{k}^{o}=0.0175$ & $d_{k}^{o}=-0.0099$ \\ \hline
\end{tabular}  
\end{center}
\end{table}

\section{Parameters for Path Interval Distribution}
For the channels in a particular Class, the path interval can be described by a Generalized Extreme Value (GEV) distribution. The PDF of GEV distribution for $i$th Class and $k$th Cluster is given in \cite{Formal_Paper} equation (10) which can be written as:
\begin{equation}
 f_{gev}\left(x;\epsilon_{i,k},\eta_{i,k},\xi_{i,k} \right) \!\! = \!\! \frac{1}{\eta_{i,k}}\!\!\left(\!\!1+\xi_{i,k}\left(\frac{x-\epsilon_{i,k}}{\eta_{i,k}} \right)\!\!  \right) ^{-\frac{1}{\xi_{i,k}}-1} \cdot \!\!\! e^{-\left(1+\xi_{i,k}\left(\frac{x-\epsilon_{i,k}}{\eta_{i,k}} \right)  \right) ^{-\frac{1}{\xi_{i,k}}}}
\end{equation} 
Parameters $\xi$ and $\eta$ in Class V should be described by power functions of cluster Index. Except for the 2 special cases in Class V, the other parameters can be written as linear functions of the cluster Index. 

\begin {table}
\caption {GEV parameters for Class V:}
\begin{center}
% use packages: array
\begin{tabular}{|l|l|}
\hline
 Expression & Fitted Result \\ \hline
$\xi_{5,k}=ak^{b}+c$ & $a=0.4063$, $b=0.2886$, $c=1.061$ \\ \hline
$\delta_{5,k}=ak^{b}+c$ &$a=1.246$, $b=0.1702$, $c=-1.892$ \\ \hline
$\mu_{5,k}=ak+b$ & $a=0.0002687$, $b=0.2033$ \\ \hline
\end{tabular}
\end{center}
\end{table}

\begin {table}
\caption {GEV parameters for Class IV:}
\begin{center}
% use packages: array
\begin{tabular}{|l|l|}
\hline
Expression & Fitted Result \\ \hline
$\xi_{4,k}=ak+b$ & $a=0.000972$, $b=2.734$ \\ \hline
$\eta_{4,k}=ak+b$ & $a=0.0009786$, $b= 0.9539$ \\ \hline
$\epsilon_{4,k}=ak+b$ & $a=0.0001653$, $b=0.3061$ \\ \hline
\end{tabular}
\end{center}
\end{table}

\begin {table}
\caption {GEV parameters for Class III:}
\begin{center}
% use packages: array
\begin{tabular}{|l|l|}
\hline
Expression & Fitted Result \\ \hline
$\xi_{3,k}=ak+b$ & $a=0.0006167$, $b=2.537$ \\ \hline
$\eta_{3,k}=ak+b$ & $a=0.0005993$, $b= 0.8095$ \\ \hline
$\epsilon_{3,k}=ak+b$ & $a=-0.00009132$, $b=0.571$ \\ \hline
\end{tabular}
\end{center}
\end{table}

\begin {table}
\caption {GEV parameters for Class II:}
\begin{center}
% use packages: array
\begin{tabular}{|l|l|}
\hline
Expression & Fitted Result \\ \hline
$\xi_{2,k}=ak+b$ & $a=0.001143$, $b=2.211$ \\ \hline
$\eta_{2,k}=ak+b$ & $a=0.0008684$, $b= 0.6979$ \\ \hline
$\epsilon_{2,k}=ak+b$ & $a=-0.00003362$, $b=0.5586$ \\ \hline
\end{tabular}
\end{center}
\end{table}

\section{Parameters for Cable Losses}
In equation (11) of \cite{Formal_Paper}, the cable loss of the powerline cable is desribed as a function of frequency and signal propagation distance. 
\begin{equation}
  A\left(f\right)=e^{-\left( a_{0}+a_{1}\cdot f^{k}\right) } e^{-jb_{0}f}
\end{equation} 
The parameters for the current cables can be written as the function of the path propagation distance:
\begin{eqnarray}
 a_{0}&=&0.0002086\cdot d + 0.0008739 \\
 a_{1}&=&0.00002644\cdot d -0.00004644 \\
 k&=&-0.00009098\cdot d + 0.8876 \\
 b_{0}&=&-0.0006432\cdot d - 0.000001126
\end{eqnarray} 
where, $d$ is the path propagation distance. The unit of $f$ is MHz.

\pagebreak

\end{document}